# ON THE LETTER FREQUENCIES AND ENTROPY
# OF WRITTEN MARATHI

JAYDEEP CHIPALKATTI

*Department of Mathematics, University of Manitoba, Winnipeg, MB R3T 2N2, Canada.*
*e-mail:* `Jaydeep.Chipalkatti@umanitoba.ca`

MIHIR KULKARNI

*Department of Astronomy, Columbia University, New York, New York 10027, U.S.A.*
*e-mail:* `mihir@astro.columbia.edu`

Abstract: We carry out a comprehensive analysis of letter frequencies in contemporary written Marathi. We determine sets of letters which statistically predominate any large generic Marathi text, and use these sets to estimate the entropy of Marathi.

## Contents

## 1. Introduction

1.1. The concept of entropy for a random variable was introduced in a foundational paper by Shannon [8]. He went on to investigate the 'entropy of printed English' in [9] and arrived at an estimate of 0.6 to 1.3 bits/letter. There have been several later attempts to calculate this value using a variety of techniques; *inter alia* by Cover and King [2], Guerrero [4], Pit-Claudel [6], and Teahan and Cleary [10]. Although these approaches do not point towards a single precise number, there is a broad consensus that it lies somewhere between 1 and 2 bits/letter.

In this paper, we attempt a similar task for written Marathi. Since its orthography is rather more intricate, one must come to terms with new complications not encountered in English. Our strategy is to find 'canonical letter sets' for Marathi, and then to *approximate* a text using these sets. It will be clear that analogous methods should be applicable to nearly all Indic languages,[1] if not more generally.

The underlying definition of entropy presupposes a specific model of a written language. Given a large generic text, we treat its string of letters as the outcome of a stationary random process, and then take the limiting value of a certain conditional entropy as the length of the string tends to infinity. Since the precise definition is somewhat technical, we begin with a series of examples which convey the intuition behind it. The canonical letter sets $\mathcal{L}_r$ will be introduced in Section 2, and the entropy calculations will be carried out in Section 3. These sets together with the probabilities assigned to their elements are given in an Appendix at the end of this article. All the programs necessary for calculations were written in Python; they can be freely downloaded from:

https://github.com/mihirskulkarni/Marathi_entropy

Readers are encouraged to write to the second named author should they need any assistance in the use of these programs.

An engaging and accessible introduction to the basic notions in information theory can be found in Applebaum [1]. We refer the reader to the treatises by Cover and Thomas [3], and Reza [7] for more comprehensive accounts of this subject.

1.2. **Entropy as the expected quantity of information.** Let us suppose that our protagonist Rasika has an appointment with a complete stranger named Robyn. At the outset, Rasika knows nothing beyond the fact that Robyn is a Dutch politician. Although Rasika does not know Robyn's gender (since the name is androgynous), she believes that Robyn is more likely to be a male (since politicians are predominantly male, even in the Netherlands). We can model this situation by a random variable $X$, which takes values in the set

---

[1]Paruchuri [5] makes an attempt to calculate the entropy of Telugu, but his method involves transliterating the source text into Roman script and working with the latter. We are sceptical of this approach.

{male, female} with respective probabilities

$$p(\text{male}) = 0.6, \quad p(\text{female}) = 0.4. \tag{1}$$

If Rasika eventually learns that Robyn is indeed a male, then the 'quantity of information' she receives[2] is $-\log_2(0.6) \approx 0.7370$, and similarly it is $-\log_2(0.4) \approx 1.3219$ if Robyn turns out to be female. In this context, the phrase 'quantity of information' has a meaning somewhat akin to 'degree of surprise'. Thus, Rasika 'receives more information' in the latter event, since *a priori* she had believed it less likely that Robyn would be a female.

In any event, before their meeting, the *expected* quantity of information for Rasika is equal to

$$-[0.6 \times \log_2(0.6) + 0.4 \times \log_2(0.4)] \approx 0.9710.$$

By definition, this is the entropy of the random variable $X$, denoted by $\mathbf{H}(X)$.

1.3. Similarly, Rasika does not know whether Robyn is tall or short, but believes that the former is likely (since the Dutch are generally tall). We model this situation by another random variable $Y$, which takes values in the set {tall, short} with respective probabilities

$$p(\text{tall}) = 0.7, \quad p(\text{short}) = 0.3. \tag{2}$$

As before,

$$\mathbf{H}(Y) = -[0.7 \times \log_2(0.7) + 0.3 \times \log_2(0.3)] \approx 0.8813.$$

Since $Y$ has a more 'uneven' distribution compared to $X$, we have $\mathbf{H}(Y) < \mathbf{H}(X)$.

1.4. **Conditional entropy.** Now our model should incorporate the fact that males have a propensity to be taller. Let us assume that the probability distribution for the pair of variables $(X, Y)$ is given by the following table:

$$\begin{aligned} p(\text{tall and male}) &= 0.5, & p(\text{short and male}) &= 0.1, \\ p(\text{tall and female}) &= 0.2, & p(\text{short and female}) &= 0.2. \end{aligned} \tag{3}$$

Notice that this is compatible with the values in (1) and (2); for instance

$$p(\text{male}) = p(\text{tall and male}) + p(\text{short and male}) = 0.5 + 0.1 = 0.6,$$

---

[2] If an event $E$ will occur with probability $p$, then the information that it has actually occurred is counted as $-\log_2(p)$ bits. This follows from a set of intuitively appealing and natural axioms first written down by Shannon (see [1, §6.6] or [7, §3-1 ff]). Henceforth, we will always use 'bit' as the unit without mentioning it explicitly.

and similarly in other cases. We have chosen the probabilities in such a way that

$$p(\text{tall and male}) = 0.5 > p(\text{tall}) \times p(\text{male}) = 0.42,$$

making $X$ and $Y$ into dependent random variables. The pair $(X, Y)$ has entropy

$$\mathbf{H}(X, Y) = -[0.5 \log_2(0.5) + \cdots + 0.2 \log_2(0.2)] \approx 1.7610,$$

which is less than $\mathbf{H}(X) + \mathbf{H}(Y) \approx 1.8522$. This reflects the fact that there is an overlap between the information conveyed by $X$ and $Y$. Now the conditional entropy

$$\mathbf{H}(Y|X) = \mathbf{H}(X, Y) - \mathbf{H}(X) \approx 0.7900$$

is a measure of the information in $Y$ which is not already implicit in $X$; or what is the same, the information in $Y$ residual to $X$.

If $X$ and $Y$ had been independent random variables (for instance, standing for gender and eye colour), then one would have $\mathbf{H}(Y|X) = \mathbf{H}(Y)$ since $X$ tells us nothing about $Y$. The lack of independence implies that $\mathbf{H}(Y|X) < \mathbf{H}(Y)$.

Broadly speaking, the same considerations apply when $X$ and $Y$ correspond to successive letters in a text.

1.5. **The entropy of the text.** Now assume that Rasika opens an English text[3], and reads a single letter at random. We can model the outcome by a random variable $X_1$ taking values in the set $\mathbb{A} = \{a, \ldots, z\}$. If all the letters were equally likely, then $\mathbf{H}(X_1) = \log_2(26) \approx 4.7004$. However, it is well-known that their probabilities are unequal; for instance, e or t occur more often than x or z. From the table in Wikipedia [11], we have the estimated probabilities

$$p(\mathsf{a}) = 0.0817, \ldots, p(\mathsf{e}) = 0.1270, \ldots, p(\mathsf{z}) = 0.0007,$$

which give the value $\mathbf{H}(X_1) \approx 4.1758$. This is called the 1-gram entropy[4] of English by Shannon [9, p. 51], and denoted by $F_1$.

Now let $X_2$ be the random variable corresponding to the next letter in the text. Again, it is well-known that $X_1$ and $X_2$ are not independent; for instance, the letter q is almost always followed by a u. In other words, the probability of a randomly chosen two-letter block being 'qu' is substantially higher than $p(\mathsf{q}) \times p(\mathsf{u})$. Hence the conditional entropy $\mathbf{H}(X_2|X_1)$ is less than $\mathbf{H}(X_2) = F_1$.

---

[3]which we treat as a continuous sequence of letters, ignoring blank spaces and punctuation.

[4]Shannon gives a value of 4.14 based upon a different probability table, but the discrepancy between the two is not worrisome.

In general, if $X_k$ denotes the random variable corresponding to the *k*-th letter, then

$$E_k = \mathbf{H}(X_k, \ldots, X_1), \tag{4}$$

is the expected quantity of information in a *k*-gram (i.e., a sequence of *k* letters). Now the *k*-gram entropy of the text is defined to be

$$F_k = \mathbf{H}(X_k | X_{k-1}, X_{k-2}, \ldots, X_1) = \mathbf{H}(X_k, X_{k-1}, \ldots, X_1) - \mathbf{H}(X_{k-1}, \ldots, X_1) = E_k - E_{k-1}. \tag{5}$$

It captures the quantity of information in $X_k$ which is not already implicit in $X_1, \ldots, X_{k-1}$. The statistical tendencies of English are such that $X_1, \ldots, X_k$ are pairwise dependent random variables, hence we have inequalities

$$F_1 > F_2 > F_3 > \ldots.$$

Finally, the 'entropy of the text' is defined to be the limit

$$F_\infty = \lim_{k \to \infty} F_k. \tag{6}$$

It is an idealisation of the information remaining in the *next unread letter*, when Rasika has already read a long way into the text. Thus, $F_\infty$ is the extent to which the next letter remains uncertain to the reader when all the previous letters are known. This is also the intuition behind the 'gambling approach to entropy' discussed at length in [2, §III] and [3, §6.6].

As mentioned above, our model assumes that the random process $(X_m)_{m \geqslant 1}$ is *stationary*, which intuitively means that the specific point at which Rasika starts reading the text is immaterial. This is of course an oversimplification, since word boundaries are an obvious source of discontinuities.[5]

1.6.  It is straightforward to estimate the $E_k$ by counting the block frequencies in a text (see Section 3.1); however, the procedure becomes computationally expensive and unreliable[6] for large *k*. This has the practical consequence that the limit in (6) is difficult to estimate with any degree of confidence.

---

[5]Notice that several English words begin with the sequence 'qu', but hardly any end with it.

[6]One salient reason behind this is that any text written by a single author will usually carry some selection bias arising from the nature of the topic and the style of the author. A legal thriller could contain several uses of an otherwise uncommon word such as 'codicil', or the author might have a tendency to overuse intensifiers such as 'very' or 'quite'. There are innumerable possibilities of this kind, any of which will distort the block frequencies. One can tentatively get around this difficulty by combining different texts into one, but it will rear its head again for larger *k*.

It is a defect of the model that the definition of $F_\infty$ has a very tenuous connection to the psychological process which lies beneath the act of reading. The average length of a word in a Marathi text is between 3 to 4 letters, and most words are smaller than 6 letters (see Section 2.2). It is reasonable to suppose that the reader usually anticipates the next unread letter *within an individual word*, but only seldom beyond it. If so, this means that our model retains at least some degree of affinity to this process for $k \leqslant 6$. However, with increasing $k$, the sequence $X_1, \ldots, X_{k-1}$ will span across not only words and sentences but entire paragraphs and chapters. It is then no longer meaningful to ask to what extent an isolated letter $X_k$ depends on the previous $X_i$, and consequently the quantity $F_k$ has little intelligible meaning outside the model.

Hence, in this paper, we only report the values of $F_k$ for $k \leqslant 6$, and make no attempt to calculate $F_\infty$. An explained in the next two sections, this involves the intermediate step of approximating a text depending on a parameter.

## 2. The Canonical Letter Sets

2.1.   The letters which can occur in an English text make a reasonably well-defined set, which remains within manageable size even if one decides to include punctuation, blank spaces and accented Roman letters.

This is not so for Marathi. It uses the *Devanagari* script, which is an abugida writing system. Since its orthography allows vowel diacritics as well as ligatures between potentially indefinite number of consonants, the number of possible letters is infinite in theory and very large in practice.[7]

However, the bulk of a generic text is usually comprised of a much smaller number of letters which occur several times. As an illustration, we present some data on a newspaper editorial, taken from the daily *Divya Marathi* (दिव्य मराठी), dated 28 April, 2014. It contains altogether 2266 letters (coming from 727 words), but only 308 distinct letters. We can order the latter by decreasing frequency of their occurrence, as in

$$(1)\ \text{त: 86 times,} \quad (2)\ \text{र: 84 times,} \quad (3)\ \text{स: 69 times,} \quad \ldots \quad (308)\ \text{कं: once.} \tag{7}$$

Then the overall pattern is that the top $n$ distinct letters in this list account for a much larger share of the text compared to $n/308$. This is shown in the left-hand plot in Figure 1, with distinct letters on the horizontal axis and the percentage share of the text on the vertical axis. For example, the point $(50, 65.66)$ is on the curve indicating that the top 50 distinct letters in (7) account for 65.66% of the text (i.e., $2266 \times 0.6566 \approx 1488$ letters).

---

[7]For example, 'श्यामची आई', a well-known fictionalised memoir by Sane Guruji, contains almost one thousand distinct letters notwithstanding its simple prose style. This number will only be greater for a book which frequently uses Sanskritisms or transliterations of English words.

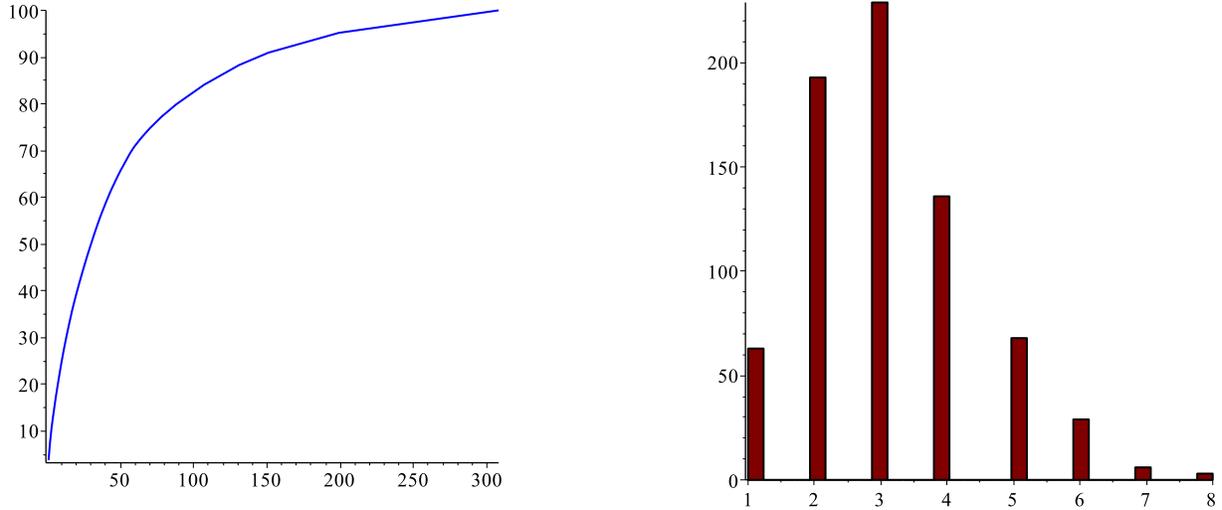

Figure 1. The text of a Marathi newspaper editorial: (1) distinct letters vs. their percentage share, and (2) distribution of word lengths

Any reasonably large Marathi text will produce a plot very similar to Figure 1. Moreover, the most frequently occurring letters are by and large common to all texts. Hence, it seems sensible to determine these letter sets once and for all.

Let $r$ denote any of the following values:

$$0.60, \quad 0.65, \quad 0.70, \quad 0.75, \quad 0.80, \quad 0.85, \quad 0.90, \quad 0.95. \tag{8}$$

We are aiming to find a 'canonical letter set' $\mathcal{L}_r$ which will account for about $r \times 100\%$ of any generic Marathi text. This turns out to be a sensible choice for the range of $r$; since such a set would not be useful for low values of $r$ (say $r = 0.40$), and would become unwieldy for values very close to 1 (say $r = 0.99$).

2.2. **Word lengths.** Although a study of word lengths in Marathi is rather extraneous to this article, it is indirectly related to the range of $k$ in the context of Sections 1.5-1.6. The distribution of word lengths for the same editorial is displayed in the right-hand plot in Figure 1, where the tallest bar shows that there are 229 words of length 3. The average word length is 3.1 for this sample, but we have found that it mostly remains within 3 and 4 for all texts and the overall shape of the bar graph remains essentially the same.

2.3. **The textual corpus.** We have chosen the *Marathi Vishwakosh* (मराठी विश्वकोश), i.e., the Marathi encyclopædia, as our corpus. It was serially published in twenty volumes during the years 1976-2015, and is now available online.[8] This choice can be justified on several grounds. Since the range of subjects in MV is very large (indeed, encyclopædic), and several hundred authors have contributed to it, there is little danger of a sample bias arising from either source. Secondly, the articles have been carefully edited to ensure that they adhere to the accepted rules of Marathi writing.[9]

In order to obliterate any potential bias arising from the alphabetical listing of entries, we have divided the entire MV into twenty 'books' $B^{(0)}, \ldots, B^{(19)}$ using the following rule. The *m*-th article in any volume of MV was put into $B^{(t)}$, where $m \equiv t \pmod{20}$. For instance, $B^{(3)}$ contains the 3rd, 23rd and in general $20i + 3$rd article from any of the volumes in MV. Each $B^{(t)}$ consists of approximately 2.2 million letters and more than three thousand distinct letters. We have used the first ten books $B^{(0)}, \ldots, B^{(9)}$ for the letter frequency calculations in this section, and reserved the rest for the entropy calculations in the next section.

2.4. **The sets $\mathcal{L}_r$.** For $t = 0, \ldots, 9$, let $\mathcal{L}_r^{(t)}$ denote the set of most frequently occurring letters[10] in $B^{(t)}$ which form just over $r \times 100\%$ of the text (exactly as in Section 2.1). Now define $\mathcal{L}_r$ to be the set of letters which occur in at least seven of the ten sets $\mathcal{L}_r^{(t)}$. This 'two-thirds majority rule' is a compromise between either taking the union $U_r = \bigcup_t \mathcal{L}_r^{(t)}$ (which would have been too permissive), or the intersection $I_r = \bigcap_t \mathcal{L}_r^{(t)}$ (which would have been too stringent).

As an illustration, the sets $U_{0.70}$ and $I_{0.70}$ have 88 and 77 elements respectively, which is a deviation of only $\pm 7\%$ from their mean. Of course, $\mathcal{L}_{0.70}$ is sandwiched between the two. In fact, it is true of all $r$ that $U_r$ and $I_r$ remain close to each other. This indicates that the $\mathcal{L}_r$ are quite robust, in the sense that their contents are not very sensitive to a variation in the defining rule.

---

[8]These are alphabetically arranged, with volume 1 ranging from अंक to आतुरचिकित्सा and volume 20 from सेई शोनागुन to ज्ञेयवाद. The complete text is available at: https://marathivishwakosh.maharashtra.gov.in/

[9]These rules were drafted by the 'Akhil Bharateeya Marathi Sahitya Mahamandal' and ratified by the Maharashtra State Government in 1962 (with a few minor changes in 1972). They have remained in general acceptance since then.

[10]We have entirely ignored punctuation, blank spaces, numerals and Roman letters. Thus, henceforth each 'letter' strictly belongs to the *Devanagari* script.

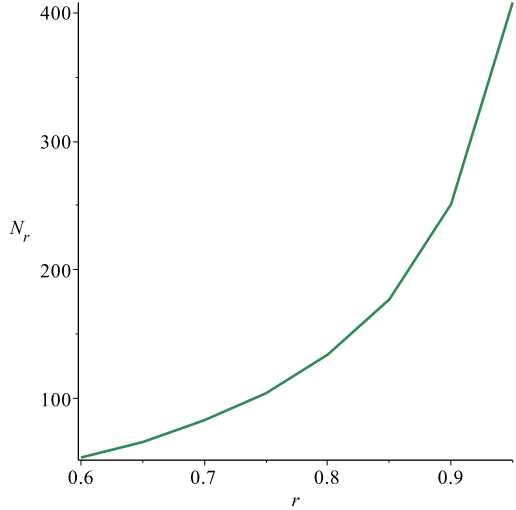

Figure 2. The change in $N_r$ with $r$

These sets are nested in the obvious way, i.e., $N_{0.60} \subset N_{0.65} \subset \cdots \subset N_{0.95}$. Their sizes $N_r = \mathrm{card}(\mathcal{L}_r)$ are:

$$N_{0.60} = 54, \quad N_{0.65} = 66, \quad N_{0.70} = 83, \quad N_{0.75} = 104, \\ N_{0.80} = 134, \quad N_{0.85} = 177, \quad N_{0.90} = 251, \quad N_{0.95} = 408. \tag{9}$$

It is easy to notice a 'law of diminishing returns' in these numbers. One needs only 54 letters to capture 60% of a text, but 80 additional letters in order to increase this share by 20%. This is even more visible from the sharply rising plot in Figure 2, which shows the change in $N_r$ with $r$.

We will represent the set $\mathcal{L}_{0.95}$ as $\{x_1, x_2, \ldots, x_{408}\}$, with the convention that $\{x_1, \ldots, x_{N_r}\} = \mathcal{L}_r$ for any smaller value of $r$.

2.5. **Letter frequencies.** As we saw earlier in Section 1.5, one can assign statistically stable probabilities to the letters a, ..., z in English. This cannot be carried out for *all* the letters in Marathi orthography, since some rare letters occur unpredictably. For instance, to the best of our knowledge, the letter 'ज्वाँ' only occurs in the word 'बूज्वाँ' indirectly borrowed from French via English. The word carries a distinctly pejorative connotation of 'middle-class complacency', especially in a phrase such as 'बूज्वाँ मनोवृत्ती' ('the bourgeois mindset'). It follows that, depending on the subject of the text and the political leanings of the author, such a letter will occur either frequently or not at all. Hence no meaningful probability can be assigned to it.

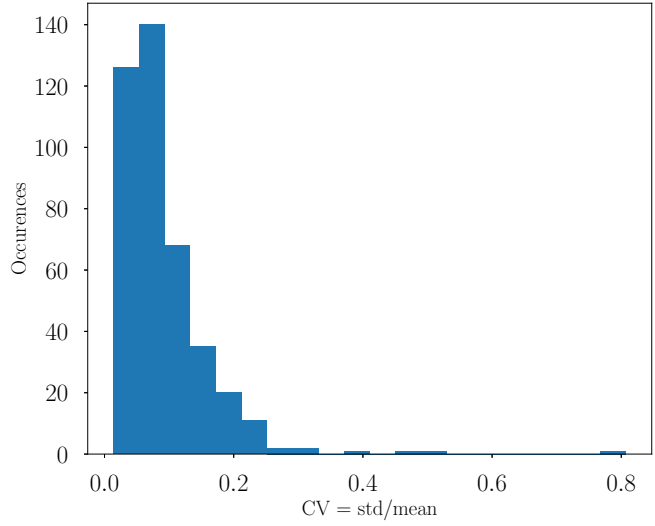

Figure 3. The bar graph for the CV of $q_i^{(0)}, \ldots, q_i^{(9)}$

However, such a project does turn out to be possible for the smaller set of letters $\mathcal{L}_{0.95} = \{x_1, \ldots, x_{408}\}$. We simply count the occurrence of each $x_i$ in each of the books $B^{(0)}, \ldots, B^{(9)}$. For $0 \leqslant t \leqslant 9$, let

$$q_i^{(t)} = \frac{\text{Number of occurrences of } x_i \text{ in the book } B^{(t)}}{\text{Total number of letters in } B^{(t)}},$$

and define $p(x_i)$ to be the arithmetic mean of the numbers $q_i^{(0)}, \ldots, q_i^{(9)}$. As seen in Figure 3, the coefficient of variation (CV) for these ten data points (defined as the ratio of their standard deviation to mean) is less than 30% for almost all $x_i$. Since the $q_i^{(t)}$ have rather low variability from book to book, we are justified in assigning the probability $p(x_i)$ to $x_i$.

The first seven letters with the largest probabilities are:

त: 0.0366, र: 0.0341, क: 0.0280, न: 0.0276, व: 0.0257, अ: 0.0205, स: 0.0203.

The rest are less that 0.02. It is noteworthy that these seven make up for 19.3%, i.e., almost one fifth of a text. The ligatures are naturally lower down in the list; the highest amongst them is 'च्या', with a probability of 0.0118, and the next highest is 'प्र' with a probability of 0.0093.

One test of correctness of the $p(x_i)$ is that the quantity $\dfrac{\sum\limits_{i=1}^{N_r} p(x_i)}{r}$ should be close to 1. We have found that it lies between 0.9886 and 0.9958 for all $r$.

A complete listing of the $\mathcal{L}_r$ together with values of $p(x_i)$ may be found at the end of this article.

2.6. **Approximating a text.** We will use these sets to approximate a given Marathi text $T$. Introduce a placeholder symbol □, and let $\alpha_r(T)$ denote the text obtained by replacing each letter not in $\mathcal{L}_r$ by □. The rationale behind this construction is that since only the infrequent letters have been replaced, the 'quantity of information' in the text should suffer only a small amount of loss. This will be borne out by the following two examples.

Given below is an instance of $\alpha_{0.75}(T_1)$:

बाजार□व□त ही □ल्याची संक□ना □र मह□ची असते. □□ म्हण□ □हकाला व□ची □ □□ □मत वाटते ती □मत. ही □मत □हकाच्या मनात असते आणि ती □र नसते. ती अनेकविध घट□च्या एक□त परिणा□□न □हकाच्या मनात साकार होते. अ□शा□□सार ए□□ व□ला □□ असण्यासाठी एक तर ती व□ □हकासाठी उप□□ असायला ह□ आणि □सरे म्हण□ ती म□दित प्रमाणात उपल□ असायला ह□. ए□ण गर□□□ जा□ प्रमाणात उपल□ असलेल्या व□ना □मत नसते □वा □चा □णाला उप□ग नाही त्या व□ला □मत नसते.

A seasoned Marathi reader would probably find this passage annoying, but should have no difficulty understanding its overall drift. It is possible to guess most of the missing letters if they are isolated, but this may well be harder if they occur consecutively.

Our second example is that of $\alpha_{0.90}(T_2)$:

तरीहि पेशवाईच्या अखेरीस असे दिसते की राजकीय वा महसूली प्रकारचा पत्रव्यवहार आणि काही बखर वा□य सोडले तर ग□ लेखनाची काही शिस्त मराठीमध्ये विकसित झाली न□ती. कोकणातली, देशावरची, विद□तली, खानदेशी अशी मराठीची जी अनेक □लसापेक्ष □लण्यातली भा□रूपे वापरात होती त्या□की प्रमाण मराठी भा□ असे कोणास म्हणावे आणि ती कशी लिहिली जावी ह्याबाबत काहीच विचार झाला न□ता. ह्याचे कारण म्हणजे लेखनवाचन ह्या गो□मध्ये काही □र□ असलेला □डितवर्ग जवळजवळ संपूर्णपणे आपले लक्ष सं□त भा□ आणि त्या भा□तील पारंपारिक ज्ञान ह्यावरच के□त करून होता. मराठी भा□चा विकास, तिची श□स□□, व्याकरण, प्रमाणलेखन ह्याकडेहि लक्ष द्यावे असे कोणाच पारंपारिक वि□नास जाणवत न□ते.

The loss of meaning in this case is very slight, and the passage is almost entirely comprehensible. We now enclose the original passages $T_1$ and $T_2$:

बाजारव्यवस्थेत ही मूल्याची संकल्पना फार महत्त्वाची असते. मूल्य म्हणजे ग्राहकाला वस्तूची जी योग्य किंमत वाटते ती किंमत. ही किंमत ग्राहकाच्या मनात असते आणि ती स्थिर नसते. ती अनेकविध घटकांच्या एकत्रित परिणामांतून ग्राहकाच्या मनात साकार होते. अर्थशास्त्रानुसार एखाद्या वस्तूला मूल्य असण्यासाठी एक तर ती वस्तू ग्राहकासाठी उपयुक्त असायला हवी आणि दुसरे म्हणजे ती मर्यादित प्रमाणात उपलब्ध असायला हवी. एकूण गरजेपेक्षा जास्त प्रमाणात उपलब्ध असलेल्या वस्तूंना किंमत नसते किंवा जिचा कुणाला उपयोग नाही त्या वस्तूला किंमत नसते.

तरीहि पेशवाईच्या अखेरीस असे दिसते की राजकीय वा महसूली प्रकारचा पत्रव्यवहार आणि काही बखर वाङ्मय सोडले तर गद्य लेखनाची काही शिस्त मराठीमध्ये विकसित झाली नव्हती. कोकणातली, देशावरची, विदर्भातली, खानदेशी अशी मराठीची जी अनेक स्थलसापेक्ष बोलण्यातली भाषारूपे वापरात होती त्यापैकी प्रमाण मराठी भाषा असे कोणास म्हणावे आणि ती कशी लिहिली जावी ह्याबाबत काहीच विचार झाला नव्हता. ह्याचे कारण म्हणजे लेखनवाचन ह्या गोष्टींमध्ये काही स्वारस्य असलेला पंडितवर्ग जवळजवळ संपूर्णपणे आपले लक्ष संस्कृत भाषा आणि त्या भाषेतील पारंपारिक ज्ञान ह्यावरच केन्द्रित करून होता. मराठी भाषेचा विकास, तिची शब्दसमृद्धि, व्याकरण, प्रमाणलेखन ह्याकडेहि लक्ष द्यावे असे कोणाच पारंपारिक विद्वानास जाणवत नव्हते.

We have carried out similar experiments with several passages and varying values of $r$. It seems safe to assert that almost all the semantic content of $T$ is retained in $\alpha_r(T)$, whenever $r \geqslant 0.85$. Notice that for a given text $T$, the proportion of missing letters in $\alpha_r(T)$ may be more or less than $1 - r$ depending on the type of vocabulary used. A passage which uses obscure Sanskritisms will lose more of its letters, whereas a children's story written in simple ligature-free words will lose fewer of them.

In the next section, we will perform all of our entropy calculations on $\alpha_r(T)$ instead of $T$ itself.

## 3. The $k$-gram Entropies

3.1. In the situation of Section 1.5, assume that our text $S$ is a sequence of $s$ letters from an alphabet $\mathbb{A}$. In particular, if $S$ is the result of applying $\alpha_r$ to a text, then the operative alphabet is $\mathbb{A} = \mathcal{L}_r \cup \{\square\}$. By a $k$-block we mean a subsequence of $k$ consecutive letters in $S$; thus we have altogether $s - k + 1$ blocks. Let $\{C_1, C_2, \dots\}$ denote the *distinct $k$-blocks*, and assume that $C_i$ occurs altogether $m_i$ times somewhere in $S$. Of course, one must have $\sum m_i = s - k + 1$. Let

$$\rho_i = \frac{m_i}{s - k + 1},$$

which is a proxy for the probability that a randomly chosen $k$-block equals $C_i$. Then we have the formula

$$E_k(S) = -\sum_i \rho_i \log_2(\rho_i).$$

By convention, $E_0(S) = 0$.

|       | r=0.60 | r=0.65 | r=0.70 | r=0.75 | r=0.80 | r=0.85 | r=0.90 | r=0.95 |
|-------|--------|--------|--------|--------|--------|--------|--------|--------|
| k=1   | 4.1523 | 4.5401 | 4.9733 | 5.3755 | 5.7870 | **6.2096** | 6.6597 | 7.1419 |
| k=2   | 3.8106 | 4.0924 | 4.3681 | 4.6114 | 4.8693 | 5.1030 | 5.3156 | 5.4579 |
| k=3   | 3.4769 | 3.6500 | 3.7901 | 3.8823 | 3.9551 | **3.9742** | 3.9247 | 3.8010 |
| k=4   | 3.0706 | 3.1138 | 3.0967 | 3.0286 | 2.9063 | **2.7189** | 2.4901 | 2.2480 |
| k=5   | 2.4812 | 2.3681 | 2.1879 | 1.9929 | 1.7620 | 1.5321 | 1.3282 | 1.1815 |
| k=6   | 1.7539 | 1.5373 | 1.2967 | 1.0948 | 0.9024 | 0.7585 | 0.6552 | 0.5937 |

Table 1. The values of $F_k^{(r)}$

3.2. As stated earlier, the books $B^{(t)}, t = 10, \ldots, 19$ had been set aside for entropy calculations. We have constructed the approximating texts $S_r^{(t)} = \alpha_r(B^{(t)})$ for all $r$-values in (8) and used the procedure above to find $E_k(S_r^{(t)})$ for $1 \leqslant k \leqslant 6$.

For each pair $(r, k)$, let $E_k^{(r)}$ denote the arithmetic mean of $E_k(S_r^{(t)})$ for $t = 10, \ldots, 19$. The CV for these ten data points is smaller than 3% in all cases, which shows that the entropy values are very stable across the books. Finally, let

$$F_k^{(r)} = E_k^{(r)} - E_{k-1}^{(r)}, \qquad \text{for } 1 \leqslant k \leqslant 6.$$

The results are given in Table 1, and the same data are plotted in Figure 4.

As remarked earlier, the value of $F_k^{(r)}$ monotonically decreases with $k$. Since the quantity of information retained in $\alpha_r(T)$ increases with $r$, it is also natural that $F_k^{(r)}$ should increase with $r$ at least for low values of $k$.

However, as $k$ moves beyond the typical word length of 4, several of the $k$-blocks tend to occur only once in the text. This is intuitively plausible, since a phrase longer than a word is less likely to appear frequently in a text compared to a word. Hence $X_k$ is entirely determined by $X_1, \ldots, X_{k-1}$ for at least those blocks, and this tends to reduce the value of $F_k^{(r)}$. However, this effect is less pronounced for small $r$, since in that case two $k$-blocks which are unequal in the text have more of a chance of becoming equal after applying $\alpha_r$. This is very likely the explanation for the fact that $F_k^{(r)}$ decreases with increasing $r$ beyond $k \geqslant 4$. One consequence of this inversion is that we have a *mesh* of lines above the region $3 \leqslant k \leqslant 4$ roughly corresponding to the average word length.

3.3. It should be mentioned in passing that we have experimented with the calculations of $F_k^{(r)}$ for other samples besides the $B^{(t)}$. Although the plot is visually similar to Figure 4 in all cases, the mesh seems to move leftward for a small sample and rightward for a large one. This is consistent with the intuition that the $k$-block selection bias is more severe for a smaller sample, which brings about an earlier onset of the inversion mentioned above. The movement of the mesh is not very pronounced in either direction, and it always remains approximately within the band $2.5 \leqslant k \leqslant 4.5$.

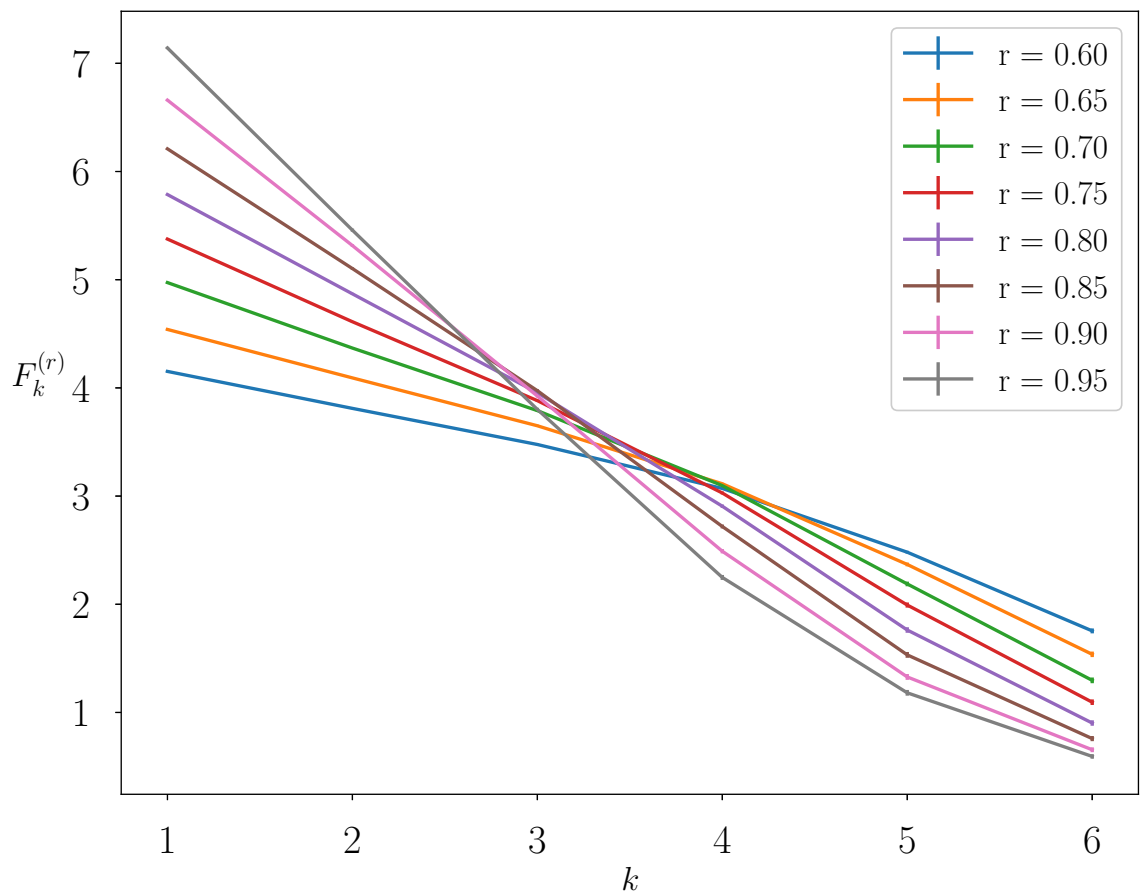

Figure 4. The behaviour of $F_k^{(r)}$

3.4. The parameter $r$ mirrors a fundamental conflict between fidelity to the original text and the reliability of our calculated result. As $r$ increases, the approximating text $\alpha_r(T)$ approaches $T$ itself. On the other hand, the size of the alphabet also increases with $r$, which renders the calculation more prone to selection bias at the level of $k$-blocks.

In light of the examples in Section 2.6, the point $r = 0.85$ is perhaps a fair compromise between these two disparate ends. Given this choice,

$$F_1^{(0.85)} = 6.2096,$$

should be seen as the Marathi analogue of the value 4.1758 for English (cf. Section 1.5). To the extent that the reader anticipates the next letter within a word, the values $F_k^{(0.85)}$ for $k = 3, 4$, aim to capture the average level of uncertainty in this anticipation. However, it cannot be emphasized too strongly that no single value can do justice to the cognitive and semantic complexities involved in the act of reading a text.

3.5. **Concluding Remarks.** Since the canonical letter sets $\mathcal{L}_r$ are stable structural features of written Marathi, they should be of interest beyond their use in calculations of entropy. It is a common theme in text compression algorithms (e.g., Huffman coding) that one should identify the most frequently occurring symbols in the source data and encode them by strings of small length. Since the elements of $\mathcal{L}_r$ play precisely this role in a Marathi text, it is natural to speculate whether they can be harnessed to a similar use.

Several Indic languages besides Marathi (such as Konkani and Hindi) use the *Devanagari* script, and those which do not (such as Bengali, Punjabi or Tamil) nevertheless use variant forms of abugida writing systems. It would be a worthwhile undertaking to see whether analogous letter sets can be found for these languages as well.

---

## Appendix: The Table of Probabilities

We enclose a listing of the letters

$$\mathcal{L}_{0.95} = \{x_1, \ldots, x_{408}\},$$

along with their assigned probabilities $p(x_i)$, as explained on page 8 of the article. For instance, the first entry means that the letter $x_1 = $ 'त' has assigned probability $p(x_1) = 3.657388 \times 10^{-2}$.

There is a dividing line after the 54th entry, 66th entry, 83rd entry and so on. These are consistent with the nesting

$$\mathcal{L}_{0.60} \subset \mathcal{L}_{0.65} \subset \cdots \subset \mathcal{L}_{0.95}.$$

Thus

$$\mathcal{L}_{0.60} = \{x_1, \ldots, x_{54}\}$$

and similarly for other values of $r$.

| $i$ | $x_i$ | unicode point sequence for $x_i$ | $p(x_i)$ |
|---|---|---|---|
| 1 | त | U+0924 | 3.657388e-02 |
| 2 | र | U+0930 | 3.413344e-02 |
| 3 | क | U+0915 | 2.803426e-02 |
| 4 | न | U+0928 | 2.760335e-02 |
| 5 | व | U+0935 | 2.572512e-02 |
| 6 | अ | U+0905 | 2.054680e-02 |
| 7 | स | U+0938 | 2.025884e-02 |
| 8 | प | U+092a | 1.692714e-02 |
| 9 | ल | U+0932 | 1.661258e-02 |
| 10 | आ | U+0906 | 1.647714e-02 |
| 11 | का | U+0915 U+093e | 1.497261e-02 |
| 12 | म | U+092e | 1.377602e-02 |
| 13 | च्या | U+091a U+094d U+092f U+093e | 1.175195e-02 |
| 14 | ता | U+0924 U+093e | 1.136662e-02 |
| 15 | ले | U+0932 U+0947 | 1.078432e-02 |
| 16 | वि | U+0935 U+093f | 1.063895e-02 |
| 17 | वा | U+0935 U+093e | 1.044438e-02 |
| 18 | ती | U+0924 U+0940 | 1.010055e-02 |
| 19 | ला | U+0932 U+093e | 1.001658e-02 |
| 20 | ना | U+0928 U+093e | 9.621307e-03 |
| 21 | सा | U+0938 U+093e | 9.589073e-03 |
| 22 | मा | U+092e U+093e | 9.513163e-03 |
| 23 | या | U+092f U+093e | 9.304941e-03 |
| 24 | प्र | U+092a U+094d U+0930 | 9.295019e-03 |
| 25 | रा | U+0930 U+093e | 9.254720e-03 |
| 26 | ग | U+0917 | 9.247231e-03 |
| 27 | य | U+092f | 9.079185e-03 |
| 28 | ते | U+0924 U+0947 | 8.888165e-03 |
| 29 | ण | U+0923 | 8.453581e-03 |
| 30 | चा | U+091a U+093e | 8.041628e-03 |
| 31 | चे | U+091a U+0947 | 7.987749e-03 |
| 32 | हे | U+0939 U+0947 | 7.545709e-03 |
| 33 | ने | U+0928 U+0947 | 7.147505e-03 |
| 34 | ज | U+091c | 6.942665e-03 |
| 35 | च | U+091a | 6.800206e-03 |
| 36 | द | U+0926 | 6.661848e-03 |
| 37 | ची | U+091a U+0940 | 6.595762e-03 |
| 38 | पा | U+092a U+093e | 6.471924e-03 |
| 39 | उ | U+0909 | 6.353647e-03 |
| 40 | ही | U+0939 U+0940 | 6.246438e-03 |
| 41 | सं | U+0938 U+0902 | 6.006461e-03 |
| 42 | नि | U+0928 U+093f | 5.926775e-03 |
| 43 | ब | U+092c | 5.865630e-03 |
| 44 | त्या | U+0924 U+094d U+092f U+093e | 5.823814e-03 |
| 45 | हो | U+0939 U+094b | 5.767016e-03 |
| 46 | ड | U+0921 | 5.718959e-03 |
| 47 | ली | U+0932 U+0940 | 5.695079e-03 |
| 48 | री | U+0930 U+0940 | 5.619677e-03 |
| 49 | णा | U+0923 U+093e | 5.441562e-03 |
| 50 | श | U+0936 | 5.369520e-03 |
| 51 | ट | U+091f | 5.287323e-03 |
| 52 | ळ | U+0933 | 5.069395e-03 |
| 53 | ह | U+0939 | 5.040355e-03 |
| 54 | हा | U+0939 U+093e | 4.999221e-03 |
| 55 | ल्या | U+0932 U+094d U+092f U+093e | 4.655049e-03 |
| 56 | के | U+0915 U+0947 | 4.509967e-03 |
| 57 | जा | U+091c U+093e | 4.412430e-03 |
| 58 | भा | U+092d U+093e | 4.408843e-03 |
| 59 | ए | U+090f | 4.187229e-03 |
| 60 | ण्या | U+0923 U+094d U+092f U+093e | 4.100178e-03 |
| 61 | इ | U+0907 | 3.974145e-03 |
| 62 | मु | U+092e U+0941 | 3.971823e-03 |
| 63 | नी | U+0928 U+0940 | 3.816529e-03 |
| 64 | शा | U+0936 U+093e | 3.814070e-03 |
| 65 | दा | U+0926 U+093e | 3.668584e-03 |
| 66 | तो | U+0924 U+094b | 3.642839e-03 |
| 67 | ये | U+092f U+0947 | 3.485317e-03 |
| 68 | सू | U+0938 U+0942 | 3.479668e-03 |
| 69 | रि | U+0930 U+093f | 3.477957e-03 |
| 70 | ध | U+0927 | 3.453459e-03 |
| 71 | णि | U+0923 U+093f | 3.450840e-03 |
| 72 | यां | U+092f U+093e U+0902 | 3.391220e-03 |
| 73 | रू | U+0930 U+0942 | 3.313950e-03 |
| 74 | से | U+0938 U+0947 | 3.307027e-03 |
| 75 | ति | U+0924 U+093f | 3.286694e-03 |
| 76 | रे | U+0930 U+0947 | 3.151157e-03 |
| 77 | णे | U+0923 U+0947 | 3.111452e-03 |
| 78 | वे | U+0935 U+0947 | 3.017336e-03 |
| 79 | त्यां | U+0924 U+094d U+092f U+093e U+0902 | 2.964957e-03 |
| 80 | गा | U+0917 U+093e | 2.942437e-03 |
| 81 | दे | U+0926 U+0947 | 2.933073e-03 |
| 82 | ख | U+0916 | 2.911625e-03 |
| 83 | मि | U+092e U+093f | 2.820485e-03 |
| 84 | शि | U+0936 U+093f | 2.811291e-03 |
| 85 | झा | U+091d U+093e | 2.754705e-03 |
| 86 | ध्ये | U+0927 U+094d U+092f U+0947 | 2.615522e-03 |
| 87 | हि | U+0939 U+093f | 2.528147e-03 |
| 88 | बा | U+092c U+093e | 2.502101e-03 |
| 89 | ळे | U+0933 U+0947 | 2.486873e-03 |
| 90 | पु | U+092a U+0941 | 2.415607e-03 |
| 91 | मी | U+092e U+0940 | 2.379287e-03 |
| 92 | त्र | U+0924 U+094d U+0930 | 2.376319e-03 |
| 93 | क्ष | U+0915 U+094d U+0937 | 2.312695e-03 |
| 94 | दि | U+0926 U+093f | 2.281731e-03 |
| 95 | ढ | U+0922 | 2.269693e-03 |
| 96 | सु | U+0938 U+0941 | 2.263429e-03 |
| 97 | की | U+0915 U+0940 | 2.257253e-03 |
| 98 | म्ह | U+092e U+094d U+0939 | 2.245389e-03 |
| 99 | शी | U+0936 U+0940 | 2.244994e-03 |
| 100 | धा | U+0927 U+093e | 2.108764e-03 |
| 101 | ष | U+0937 | 2.080513e-03 |
| 102 | रो | U+0930 U+094b | 2.078140e-03 |
| 103 | ठी | U+0920 U+0940 | 2.077803e-03 |
| 104 | घ | U+0918 | 2.051672e-03 |
| 105 | को | U+0915 U+094b | 2.030955e-03 |
| 106 | पू | U+092a U+0942 | 1.890996e-03 |
| 107 | णी | U+0923 U+0940 | 1.878765e-03 |
| 108 | लि | U+0932 U+093f | 1.859826e-03 |
| 109 | खा | U+0916 U+093e | 1.808284e-03 |
| 110 | ऊ | U+090a | 1.771184e-03 |
| 111 | ळा | U+0933 U+093e | 1.752087e-03 |
| 112 | च्या | U+0931 U+094d U+092f U+093e | 1.723443e-03 |

| 113 | रं | U+0930 U+0902 | 1.712288e-03 |
| --- | --- | --- | --- |
| 114 | यो | U+092f U+094b | 1.695707e-03 |
| 115 | भ | U+092d | 1.679965e-03 |
| 116 | कां | U+0915 U+093e U+0902 | 1.621287e-03 |
| 117 | सि | U+0938 U+093f | 1.610573e-03 |
| 118 | तू | U+0924 U+0942 | 1.578732e-03 |
| 119 | व्या | U+0935 U+094d U+092f U+093e | 1.573697e-03 |
| 120 | प्रा | U+092a U+094d U+0930 U+093e | 1.557280e-03 |
| 121 | वी | U+0935 U+0940 | 1.541094e-03 |
| 122 | दी | U+0926 U+0940 | 1.533521e-03 |
| 123 | व्य | U+0935 U+094d U+092f | 1.531636e-03 |
| 124 | यु | U+092f U+0941 | 1.526110e-03 |
| 125 | द्ध | U+0926 U+094d U+0927 | 1.520075e-03 |
| 126 | जे | U+091c U+0947 | 1.507464e-03 |
| 127 | धी | U+0927 U+0940 | 1.502589e-03 |
| 128 | मे | U+092e U+0947 | 1.484242e-03 |
| 129 | मो | U+092e U+094b | 1.480014e-03 |
| 130 | फ | U+092b | 1.443253e-03 |
| 131 | थ | U+0925 | 1.442163e-03 |
| 132 | स्था | U+0938 U+094d U+0925 U+093e | 1.437526e-03 |
| 133 | कि | U+0915 U+093f | 1.435810e-03 |
| 134 | र्व | U+0930 U+094d U+0935 | 1.425979e-03 |
| 135 | ई | U+0908 | 1.411442e-03 |
| 136 | जी | U+091c U+0940 | 1.398205e-03 |
| 137 | डा | U+0921 U+093e | 1.396542e-03 |
| 138 | लो | U+0932 U+094b | 1.382865e-03 |
| 139 | डे | U+0921 U+0947 | 1.359452e-03 |
| 140 | चि | U+091a U+093f | 1.356059e-03 |
| 141 | डी | U+0921 U+0940 | 1.306748e-03 |
| 142 | किं | U+0915 U+093f U+0902 | 1.292923e-03 |
| 143 | दो | U+0926 U+094b | 1.286898e-03 |
| 144 | ठ | U+0920 | 1.251411e-03 |
| 145 | बं | U+092c U+0902 | 1.251270e-03 |
| 146 | णू | U+0923 U+0942 | 1.229206e-03 |
| 147 | द्र | U+0926 U+094d U+0930 | 1.227272e-03 |
| 148 | शे | U+0936 U+0947 | 1.226076e-03 |
| 149 | टा | U+091f U+093e | 1.197967e-03 |
| 150 | नु | U+0928 U+0941 | 1.186060e-03 |
| 151 | धि | U+0927 U+093f | 1.174406e-03 |
| 152 | ळी | U+0933 U+0940 | 1.157310e-03 |
| 153 | गो | U+0917 U+094b | 1.151472e-03 |
| 154 | स्व | U+0938 U+094d U+0935 | 1.145167e-03 |
| 155 | अं | U+0905 U+0902 | 1.131263e-03 |
| 156 | मू | U+092e U+0942 | 1.107721e-03 |
| 157 | तु | U+0924 U+0941 | 1.092187e-03 |
| 158 | पे | U+092a U+0947 | 1.090012e-03 |
| 159 | गु | U+0917 U+0941 | 1.077573e-03 |
| 160 | रां | U+0930 U+093e U+0902 | 1.074798e-03 |
| 161 | नं | U+0928 U+0902 | 1.057957e-03 |
| 162 | थे | U+0925 U+0947 | 1.041376e-03 |
| 163 | क्त | U+0915 U+094d U+0924 | 1.028358e-03 |
| 164 | ज्ञा | U+091c U+094d U+091e U+093e | 1.023034e-03 |
| 165 | रु | U+0930 U+0941 | 1.020169e-03 |
| 166 | र्ण | U+0930 U+094d U+0923 | 1.015437e-03 |
| 167 | पि | U+092a U+093f | 1.010738e-03 |
| 168 | भू | U+092d U+0942 | 1.003810e-03 |
| 169 | स्त | U+0938 U+094d U+0924 | 1.003810e-03 |
| 170 | द्या | U+0926 U+094d U+092f U+093e | 1.003401e-03 |
| 171 | गी | U+0917 U+0940 | 9.995399e-04 |
| 172 | गे | U+0917 U+0947 | 9.924893e-04 |
| 173 | दु | U+0926 U+0941 | 9.693678e-04 |
| 174 | ऑ | U+0911 | 9.634635e-04 |
| 175 | क्षा | U+0915 U+094d U+0937 U+093e | 9.513745e-04 |
| 176 | कृ | U+0915 U+0943 | 9.201014e-04 |
| 177 | टी | U+091f U+0940 | 9.172285e-04 |
| 178 | र्य | U+0930 U+094d U+092f | 9.137319e-04 |
| 179 | ओ | U+0913 | 9.135991e-04 |
| 180 | चे | U+0918 U+0947 | 9.067191e-04 |
| 181 | क्रि | U+0915 U+094d U+0930 U+093f | 8.935280e-04 |
| 182 | ष्ट | U+0937 U+094d U+091f | 8.906519e-04 |
| 183 | ह्या | U+0939 U+094d U+092f U+093e | 8.871830e-04 |
| 184 | टि | U+091f U+093f | 8.571785e-04 |
| 185 | नां | U+0928 U+093e U+0902 | 8.536182e-04 |
| 186 | जि | U+091c U+093f | 8.516657e-04 |
| 187 | तं | U+0924 U+0902 | 8.463402e-04 |
| 188 | स्प | U+0938 U+094d U+092a | 8.285075e-04 |
| 189 | सो | U+0938 U+094b | 8.238033e-04 |
| 190 | ख्या | U+0916 U+094d U+092f U+093e | 8.235977e-04 |
| 191 | त्त | U+0924 U+094d U+0924 | 8.022542e-04 |
| 192 | न्य | U+0928 U+094d U+092f | 8.022079e-04 |
| 193 | शो | U+0936 U+094b | 7.938889e-04 |
| 194 | इं | U+0907 U+0902 | 7.897199e-04 |
| 195 | था | U+0925 U+093e | 7.744666e-04 |
| 196 | पो | U+092a U+094b | 7.684973e-04 |
| 197 | लां | U+0932 U+093e U+0902 | 7.650179e-04 |
| 198 | पी | U+092a U+0940 | 7.641254e-04 |
| 199 | कु | U+0915 U+0941 | 7.362792e-04 |
| 200 | र्थ | U+0930 U+094d U+0925 | 7.301794e-04 |
| 201 | डि | U+0921 U+093f | 7.255221e-04 |
| 202 | मं | U+092e U+0902 | 7.182698e-04 |
| 203 | क्षे | U+0915 U+094d U+0937 U+0947 | 7.114714e-04 |
| 204 | क्र | U+0915 U+094d U+0930 | 7.101724e-04 |
| 205 | त्रा | U+0924 U+094d U+0930 U+093e | 7.055124e-04 |
| 206 | त्य | U+0924 U+094d U+092f | 6.957022e-04 |
| 207 | र्मा | U+0930 U+094d U+092e U+093e | 6.928328e-04 |
| 208 | भि | U+092d U+093f | 6.907657e-04 |
| 209 | फा | U+092b U+093e | 6.864620e-04 |
| 210 | बी | U+092c U+0940 | 6.829675e-04 |
| 211 | ध्य | U+0927 U+094d U+092f | 6.813111e-04 |
| 212 | ठे | U+0920 U+0947 | 6.629373e-04 |
| 213 | मां | U+092e U+093e U+0902 | 6.579398e-04 |
| 214 | खे | U+0916 U+0947 | 6.542353e-04 |
| 215 | ग्र | U+0917 U+094d U+0930 | 6.489028e-04 |
| 216 | यं | U+092f U+0902 | 6.465844e-04 |
| 217 | ठा | U+0920 U+093e | 6.375066e-04 |
| 218 | त्रि | U+0924 U+094d U+0930 U+093f | 6.315878e-04 |
| 219 | घा | U+0918 U+093e | 6.245179e-04 |
| 220 | दू | U+0926 U+0942 | 6.203794e-04 |
| 221 | वै | U+0935 U+0948 | 6.195761e-04 |
| 222 | र्म | U+0930 U+094d U+092e | 6.193139e-04 |
| 223 | टे | U+091f U+0947 | 6.164124e-04 |
| 224 | डू | U+0921 U+0942 | 6.129222e-04 |
| 225 | ल्या | U+0933 U+094d U+092f U+093e | 6.081229e-04 |
| 226 | स्थि | U+0938 U+094d U+0925 U+093f | 6.067235e-04 |
| 227 | सी | U+0938 U+0940 | 6.046919e-04 |
| 228 | बे | U+092c U+0947 | 6.023578e-04 |
| 229 | र्या | U+0930 U+094d U+092f U+093e | 6.008597e-04 |
| 230 | सां | U+0938 U+093e U+0902 | 6.007119e-04 |
| 231 | र्ग | U+0930 U+094d U+0917 | 5.985771e-04 |
| 232 | वृ | U+0935 U+0943 | 5.975661e-04 |
| 233 | कू | U+0915 U+0942 | 5.968740e-04 |
| 234 | त् | U+0924 U+094d | 5.816545e-04 |
| 235 | दृ | U+0926 U+0943 | 5.770472e-04 |
| 236 | स्त्र | U+0938 U+094d U+0924 U+094d U+0930 | 5.751610e-04 |
| 237 | जो | U+091c U+094b | 5.739565e-04 |
| 238 | तः | U+0924 U+0903 | 5.737528e-04 |
| 239 | ज्या | U+091c U+094d U+092f U+093e | 5.709583e-04 |
| 240 | गि | U+0917 U+093f | 5.621468e-04 |
| 241 | र्श | U+0930 U+094d U+0936 | 5.586509e-04 |
| 242 | गां | U+0917 U+093e U+0902 | 5.563543e-04 |
| 243 | र्यं | U+0930 U+094d U+092f U+0902 | 5.562485e-04 |
| 244 | शां | U+0936 U+093e U+0902 | 5.477564e-04 |
| 245 | व्हा | U+0935 U+094d U+0939 U+093e | 5.454220e-04 |
| 246 | थी | U+0925 U+0940 | 5.431162e-04 |
| 247 | न्न | U+0928 U+094d U+0928 | 5.416427e-04 |
| 248 | लं | U+0932 U+0902 | 5.358638e-04 |
| 249 | हु | U+0939 U+0941 | 5.244398e-04 |
| 250 | तां | U+0924 U+093e U+0902 | 5.240902e-04 |
| 251 | द्यु | U+0926 U+094d U+092f U+0941 | 5.180537e-04 |
| 252 | द्रा | U+0926 U+094d U+0930 U+093e | 5.170950e-04 |

| 253 | यू | U+092f U+0942 | 5.134925e-04 |
|---|---|---|---|
| 254 | पृ | U+092a U+0943 | 5.112637e-04 |
| 255 | ल्प | U+0932 U+094d U+092a | 5.095946e-04 |
| 256 | ग्रं | U+0917 U+094d U+0930 U+0902 | 5.095733e-04 |
| 257 | न्या | U+0928 U+094d U+092f U+093e | 5.078778e-04 |
| 258 | वू | U+0935 U+0942 | 5.012532e-04 |
| 259 | कं | U+0915 U+0902 | 4.835992e-04 |
| 260 | क्ती | U+0915 U+094d U+0924 U+0940 | 4.829771e-04 |
| 261 | ळू | U+0933 U+0942 | 4.790124e-04 |
| 262 | द्यो | U+0926 U+094d U+092f U+094b | 4.786634e-04 |
| 263 | त्पा | U+0924 U+094d U+092a U+093e | 4.753833e-04 |
| 264 | नो | U+0928 U+094b | 4.716525e-04 |
| 265 | यी | U+092f U+0940 | 4.715160e-04 |
| 266 | ड्या | U+0921 U+094d U+092f U+093e | 4.706225e-04 |
| 267 | ढे | U+0922 U+0947 | 4.688745e-04 |
| 268 | तीं | U+0924 U+0940 U+0902 | 4.611397e-04 |
| 269 | कें | U+0915 U+0947 U+0902 | 4.610359e-04 |
| 270 | क्षि | U+0915 U+094d U+0937 U+093f | 4.593169e-04 |
| 271 | बि | U+092c U+093f | 4.565454e-04 |
| 272 | जू | U+091c U+0942 | 4.557565e-04 |
| 273 | णां | U+0923 U+093e U+0902 | 4.556220e-04 |
| 274 | पै | U+092a U+0948 | 4.536566e-04 |
| 275 | पं | U+092a U+0902 | 4.524576e-04 |
| 276 | बो | U+092c U+094b | 4.506822e-04 |
| 277 | त्त्वा | U+0924 U+094d U+0924 U+094d U+0935 U+093e | 4.410859e-04 |
| 278 | वं | U+0935 U+0902 | 4.356101e-04 |
| 279 | झ | U+091d | 4.260437e-04 |
| 280 | र्मि | U+0930 U+094d U+092e U+093f | 4.252365e-04 |
| 281 | खं | U+0916 U+0902 | 4.222682e-04 |
| 282 | ष्टी | U+0937 U+094d U+091f U+0940 | 4.206081e-04 |
| 283 | षा | U+0937 U+093e | 4.159934e-04 |
| 284 | ढी | U+0922 U+0940 | 4.141359e-04 |
| 285 | टो | U+091f U+094b | 4.118945e-04 |
| 286 | खी | U+0916 U+0940 | 4.063354e-04 |
| 287 | र्त | U+0930 U+094d U+0924 | 3.966103e-04 |

| 288 | ष्ठ | U+0937 U+094d U+0920 | 3.956403e-04 |
| --- | --- | --- | --- |
| 289 | फि | U+092b U+093f | 3.936879e-04 |
| 290 | त्ये | U+0924 U+094d U+092f U+0947 | 3.915053e-04 |
| 291 | त्त्व | U+0924 U+094d U+0924 U+094d U+0935 | 3.898629e-04 |
| 292 | पां | U+092a U+093e U+0902 | 3.868452e-04 |
| 293 | बां | U+092c U+093e U+0902 | 3.827041e-04 |
| 294 | कॅ | U+0915 U+0945 | 3.818322e-04 |
| 295 | धू | U+0927 U+0942 | 3.783440e-04 |
| 296 | ठि | U+0920 U+093f | 3.767786e-04 |
| 297 | व्ह | U+0935 U+094d U+0939 | 3.762222e-04 |
| 298 | श्र | U+0936 U+094d U+0930 | 3.761525e-04 |
| 299 | ख्य | U+0916 U+094d U+092f | 3.732798e-04 |
| 300 | प्रे | U+092a U+094d U+0930 U+0947 | 3.720216e-04 |
| 301 | प्त | U+092a U+094d U+0924 | 3.696942e-04 |
| 302 | खो | U+0916 U+094b | 3.650234e-04 |
| 303 | श्चि | U+0936 U+094d U+091a U+093f | 3.593230e-04 |
| 304 | त्व | U+0924 U+094d U+0935 | 3.587317e-04 |
| 305 | हू | U+0939 U+0942 | 3.578619e-04 |
| 306 | औ | U+0914 | 3.575919e-04 |
| 307 | स्ट | U+0938 U+094d U+091f | 3.558441e-04 |
| 308 | त्म | U+0924 U+094d U+092e | 3.546704e-04 |
| 309 | शु | U+0936 U+0941 | 3.509728e-04 |
| 310 | द्य | U+0926 U+094d U+092f | 3.506052e-04 |
| 311 | ध्या | U+0927 U+094d U+092f U+093e | 3.459948e-04 |
| 312 | श्य | U+0936 U+094d U+092f | 3.449968e-04 |
| 313 | लू | U+0932 U+0942 | 3.445081e-04 |
| 314 | त्ती | U+0924 U+094d U+0924 U+0940 | 3.437925e-04 |
| 315 | भ्या | U+092d U+094d U+092f U+093e | 3.426495e-04 |
| 316 | दं | U+0926 U+0902 | 3.413673e-04 |
| 317 | स्थे | U+0938 U+094d U+0925 U+0947 | 3.391332e-04 |
| 318 | डो | U+0921 U+094b | 3.374135e-04 |
| 319 | मृ | U+092e U+0943 | 3.372865e-04 |
| 320 | जु | U+091c U+0941 | 3.351072e-04 |
| 321 | आं | U+0906 U+0902 | 3.342616e-04 |
| 322 | उं | U+0909 U+0902 | 3.339408e-04 |
| 323 | ष्क | U+0937 U+094d U+0915 | 3.316084e-04 |
| 324 | ज्य | U+091c U+094d U+092f | 3.246780e-04 |
| 325 | न्स | U+0928 U+094d U+0938 | 3.241711e-04 |
| 326 | र्ष | U+0930 U+094d U+0937 | 3.225897e-04 |
| 327 | त्रे | U+0924 U+094d U+0930 U+0947 | 3.213323e-04 |
| 328 | फे | U+092b U+0947 | 3.207459e-04 |
| 329 | फु | U+092b U+0941 | 3.164002e-04 |
| 330 | त्ता | U+0924 U+094d U+0924 U+093e | 3.154381e-04 |
| 331 | बु | U+092c U+0941 | 3.152924e-04 |
| 332 | स्वा | U+0938 U+094d U+0935 U+093e | 3.148166e-04 |
| 333 | ठ्या | U+0920 U+094d U+092f U+093e | 3.130627e-04 |
| 334 | कॉ | U+0915 U+0949 | 3.101536e-04 |
| 335 | धु | U+0927 U+0941 | 3.005843e-04 |
| 336 | सें | U+0938 U+0947 U+0902 | 3.002396e-04 |
| 337 | र्गा | U+0930 U+094d U+0917 U+093e | 2.988736e-04 |
| 338 | ष्ण | U+0937 U+094d U+0923 | 2.987154e-04 |
| 339 | भे | U+092d U+0947 | 2.932702e-04 |
| 340 | स्तू | U+0938 U+094d U+0924 U+0942 | 2.914441e-04 |
| 341 | र्वी | U+0930 U+094d U+0935 U+0940 | 2.817926e-04 |
| 342 | श्व | U+0936 U+094d U+0935 | 2.805251e-04 |
| 343 | भो | U+092d U+094b | 2.802767e-04 |
| 344 | क्य | U+0915 U+094d U+092f | 2.791464e-04 |
| 345 | स्ता | U+0938 U+094d U+0924 U+093e | 2.790282e-04 |
| 346 | द्धा | U+0926 U+094d U+0927 U+093e | 2.760233e-04 |
| 347 | क्ष्म | U+0915 U+094d U+0937 U+094d U+092e | 2.735249e-04 |
| 348 | द्वा | U+0926 U+094d U+0935 U+093e | 2.725872e-04 |
| 349 | चां | U+091a U+093e U+0902 | 2.688451e-04 |
| 350 | चं | U+091a U+0902 | 2.635774e-04 |
| 351 | त्प | U+0924 U+094d U+092a | 2.622522e-04 |
| 352 | बू | U+092c U+0942 | 2.621631e-04 |
| 353 | गं | U+0917 U+0902 | 2.611373e-04 |
| 354 | स्कृ | U+0938 U+094d U+0915 U+0943 | 2.609832e-04 |
| 355 | ल्य | U+0932 U+094d U+092f | 2.589700e-04 |
| 356 | चौ | U+091a U+094c | 2.564296e-04 |
| 357 | ष्ट्री | U+0937 U+094d U+091f U+094d U+0930 U+0940 | 2.560307e-04 |
| 358 | कं | U+0930 U+094d U+0915 | 2.557724e-04 |
| 359 | र्च | U+0930 U+094d U+091a | 2.553882e-04 |
| 360 | ॲ | U+0972 | 2.551663e-04 |
| 361 | ग्य | U+0917 U+094d U+092f | 2.547179e-04 |
| 362 | वां | U+0935 U+093e U+0902 | 2.541402e-04 |
| 363 | थो | U+0925 U+094b | 2.520394e-04 |
| 364 | टां | U+091f U+093e U+0902 | 2.473285e-04 |
| 365 | स्त्री | U+0938 U+094d U+0924 U+094d U+0930 U+0940 | 2.437484e-04 |
| 366 | हिं | U+0939 U+093f U+0902 | 2.426824e-04 |
| 367 | थि | U+0925 U+093f | 2.397451e-04 |
| 368 | ण्यां | U+0923 U+094d U+092f U+093e U+0902 | 2.396047e-04 |
| 369 | षि | U+0937 U+093f | 2.392504e-04 |
| 370 | जां | U+091c U+093e U+0902 | 2.385421e-04 |
| 371 | व्हि | U+0935 U+094d U+0939 U+093f | 2.379549e-04 |
| 372 | षे | U+0937 U+0947 | 2.363632e-04 |
| 373 | क्या | U+0915 U+094d U+092f U+093e | 2.361620e-04 |
| 374 | मॅ | U+092e U+0945 | 2.352250e-04 |
| 375 | त्वा | U+0924 U+094d U+0935 U+093e | 2.351950e-04 |
| 376 | त्रां | U+0924 U+094d U+0930 U+093e U+0902 | 2.338291e-04 |
| 377 | ग्री | U+0917 U+094d U+0930 U+0940 | 2.332148e-04 |
| 378 | च्च | U+091a U+094d U+091a | 2.329602e-04 |
| 379 | च्यां | U+0931 U+094d U+092f U+093e U+0902 | 2.315615e-04 |
| 380 | ल्ले | U+0932 U+094d U+0932 U+0947 | 2.311707e-04 |
| 381 | ब्रि | U+092c U+094d U+0930 U+093f | 2.310448e-04 |
| 382 | न्ही | U+0928 U+094d U+0939 U+0940 | 2.298708e-04 |
| 383 | ग्रा | U+0917 U+094d U+0930 U+093e | 2.293336e-04 |
| 384 | अँ | U+0905 U+0901 | 2.286675e-04 |
| 385 | र्वां | U+0930 U+094d U+0935 U+093e U+0902 | 2.265347e-04 |
| 386 | ह्यां | U+0939 U+094d U+092f U+093e U+0902 | 2.226971e-04 |
| 387 | र्था | U+0930 U+094d U+0925 U+093e | 2.222954e-04 |
| 388 | णु | U+0923 U+0941 | 2.206283e-04 |
| 389 | ढा | U+0922 U+093e | 2.201793e-04 |
| 390 | शै | U+0936 U+0948 | 2.186492e-04 |
| 391 | क्ति | U+0915 U+094d U+0924 U+093f | 2.184093e-04 |
| 392 | ऐ | U+0910 | 2.143018e-04 |

| 393 | त्स | U+0924 U+094d U+0938 | 2.139978e-04 |
|---|---|---|---|
| 394 | स्वी | U+0938 U+094d U+0935 U+0940 | 2.139558e-04 |
| 395 | र्भ | U+0930 U+094d U+092d | 2.124561e-04 |
| 396 | षां | U+0937 U+093e U+0902 | 2.120543e-04 |
| 397 | स्त्रा | U+0938 U+094d U+0924 U+094d U+0930 U+093e | 2.110849e-04 |
| 398 | नै | U+0928 U+0948 | 2.107411e-04 |
| 399 | र्जा | U+0930 U+094d U+091c U+093e | 2.105189e-04 |
| 400 | जं | U+091c U+0902 | 2.101618e-04 |
| 401 | सिं | U+0938 U+093f U+0902 | 2.100975e-04 |
| 402 | न्हा | U+0928 U+094d U+0939 U+093e | 2.067588e-04 |
| 403 | ब्ध | U+092c U+094d U+0927 | 2.029806e-04 |
| 404 | बिं | U+092c U+093f U+0902 | 2.002168e-04 |
| 405 | द्द | U+0926 U+094d U+0926 | 1.989715e-04 |
| 406 | स्ती | U+0938 U+094d U+0924 U+0940 | 1.921736e-04 |
| 407 | द्धां | U+0926 U+094d U+0927 U+093e U+0902 | 1.897260e-04 |
| 408 | धो | U+0927 U+094b | 1.892809e-04 |

28